\documentclass[fleqn,usenatbib]{mnras}

\usepackage[T1]{fontenc}
\usepackage{amsmath}
\DeclareRobustCommand{\VAN}[3]{#2}
\let\VANthebibliography\thebibliography
\def\thebibliography{\DeclareRobustCommand{\VAN}[3]{##3}\VANthebibliography}

\usepackage{graphicx}	% Including figure files
\usepackage{amsmath}	% Advanced maths commands
\usepackage{amssymb}	% Extra maths symbols
\usepackage{txfonts}
\usepackage{color}
\usepackage{ulem}
\usepackage{caption}
\usepackage{multirow}

\newcommand\Tstrut{\rule{0pt}{2.6ex}}
\newcommand\Bstrut{\rule[-0.9ex]{0pt}{0pt}}

\title[Interstellar comet 2I/Borisov]{Activity of the first interstellar comet 2I/Borisov around perihelion: Results from Indian observatories}
\author[Aravind K. et al.,] {Aravind Krishnakumar$^{1,2}$\thanks{E-mail: aravindk@prl.res.in, aravind139@gmail.com},
Shashikiran Ganesh$^{1}$,
Kumar Venkataramani$^{3}$,
\newauthor
Devendra Sahu,$^{4}$
Dorje Angchuk,$^{4}$
Thirupathi Sivarani,$^{4}$
Athira Unni$^{4}$\\
$^{1}$Physical Research Laboratory, Ahmedabad, India\\
$^{2}$Institute of Technology Gandhinagar, Gandhinagar, India\\
$^{3}$Auburn University, Auburn, USA \\
$^{4}$Indian Institute of Astrophysics, Bangalore, India
}

\date{Accepted XXX. Received YYY; in original form ZZZ}

\pubyear{2021}

\begin{document}
\label{firstpage}
\pagerange{\pageref{firstpage}--\pageref{lastpage}}
\maketitle

\begin{abstract}
   Comet 2I/Borisov is the first true interstellar comet discovered.
   Here we present results from observational programs at two Indian observatories, 2~m Himalayan Chandra Telescope at the Indian Astronomical Observatory, Hanle (HCT) and 1.2~m telescope at the Mount Abu Infrared Observatory (MIRO). 
   Two epochs of imaging and spectroscopy were carried out at the HCT and three epochs of imaging at MIRO.  We found CN to be the dominant molecular emission on both epochs, 31/11/2019 and 22/12/2019, at distances of r$_H$ = 2.013 and 2.031 AU respectively.  The comet was inferred to be relatively depleted in Carbon bearing molecules on the basis of low $C_2$ and $C_3$ abundances. We find the production rate ratio, Q(C$_2$)/Q(CN) = 0.54 $\pm$ 0.18, pre-perihelion and Q(C$_2$)/Q(CN) = 0.34 $\pm$ 0.12 post-perihelion. 
   This classifies the comet as being moderately depleted in carbon chain molecules.  Using the results from spectroscopic observations, we believe the comet to have a chemically heterogeneous surface having variation in abundance of carbon chain molecules.  From imaging observations we infer a dust-to-gas ratio similar to carbon chain depleted comets of the Solar system.  We also compute the nucleus size to be in the range $0.18\leq r \leq 3.1$ Km. Our observations show that 2I/Borisov's behaviour is analogous to that of the Solar system comets.
\end{abstract}


%%%%%%%%%%%%%%%%%%%%%%%%%%%%%%%%%%%%%%%%%%%%%%%%%%

%%%%%%%%%%%%%%%%% BODY OF PAPER %%%%%%%%%%%%%%%%%%

\section{Introduction}
  Comets are made up of pristine material inherited from the proto-solar nebula and they have spent most of their time in the farther reaches of the Solar system. In consequence, they can be considered to be the time-capsules of the early Solar system. Hence, studying the various aspects of these minor bodies can help us gain insight into the conditions that prevailed during the formation of the Solar system. In this aspect, comparing comets from our Solar system with interstellar ones can shed light on the difference/similarity in the materials present in different proto-stellar systems. Even after centuries of comet observations and decades after the initial prediction by \cite{sen&rana}, no one had observed an interstellar object, passing through the inner Solar system, until October 2017 when ‘Oumuamua (1I/2017 U1) was discovered. Even though there was non-gravitational acceleration found in its orbit, around perihelion, which is usually a result of outgassing \citep[eg.,][]{micheli_oumuamua}, ‘Oumuamua is observed to be completely asteroidal in nature \citep{i1_asteroid,oumuamua_Jewitt}. This behaviour limited the observation of the object due to its faintness. Later on 30 August 2019, Gennady Borisov using his self-built 0.65~m telescope discovered a comet like body.  This was later identified to be the first ever interstellar comet to be observed passing through the Solar system. The comet possessed a very large eccentricity of \textit{e = 3.379} and a very high hyperbolic excess velocity of v $\sim$ 32 Km/s \citep{guzik_borisov}, further confirming the interstellar origin. The interstellar comet, initially identified as C/2019 Q4, was later named 2I/Borisov\footnote{https://minorplanetcenter.net/mpec/K19/K19S72.html} by IAU. \cite{CN_detected} were the first to report the detection of CN in the interstellar comet. Later, \cite{leon_borisov}, \cite{opitom_borisov} and \cite{kareta_borisov}, have all reported the clear detection of CN along with an upper limit to the production rate of C$_2$(0-0) emissions. \cite{lin_borisov} and \cite{borisov_muse} have reported the clear detection of both CN and C$_2$ in their spectrum, with the latter work having the most detailed spectrum of 2I/Borisov reporting the detection of well-resolved C$_2$, NH$_2$ and CN emissions.\\
   In this paper, we discuss the spectroscopic and imaging observations carried out from two Indian observatories during November and December 2019, to study the evolution of molecular emissions and also to put constraints on the physical characteristics of the rare interstellar comet 2I/Borisov.   Section \ref{sec:1} describes the observations from both observatories.  Section \ref{sec:2} discusses the data reduction and analysis methods used.  Finally, we discuss the spectroscopic and imaging results in section \ref{sec:3}.  

\section{Observations}
\label{sec:1}
Observations of the interstellar comet, 2I/Borisov, were carried out using two Indian observatories, the 2~m Himalayan Chandra Telescope(HCT) operated by the Indian Institute of Astrophysics at Hanle, Ladakh and the 1.2~m telescope at Mount Abu InfraRed Observatory (MIRO) operated by the Physical Research Laboratory at Mount Abu, Rajasthan. 

In the following sub-sections we describe, briefly, the details of the observations.
The observational log, including the heliocentric distance, geocentric distance, phase angle and airmass at the time of observations are as given in Table \ref{observations_1}.   The NASA JPL HORIZONS\footnote{\url{https://ssd.jpl.nasa.gov/horizons.cgi}} service was used to generate the ephemerides for the comet for both the observing locations. \\

   \begin{figure*}
   \centering
   \includegraphics[width=0.9\textwidth]{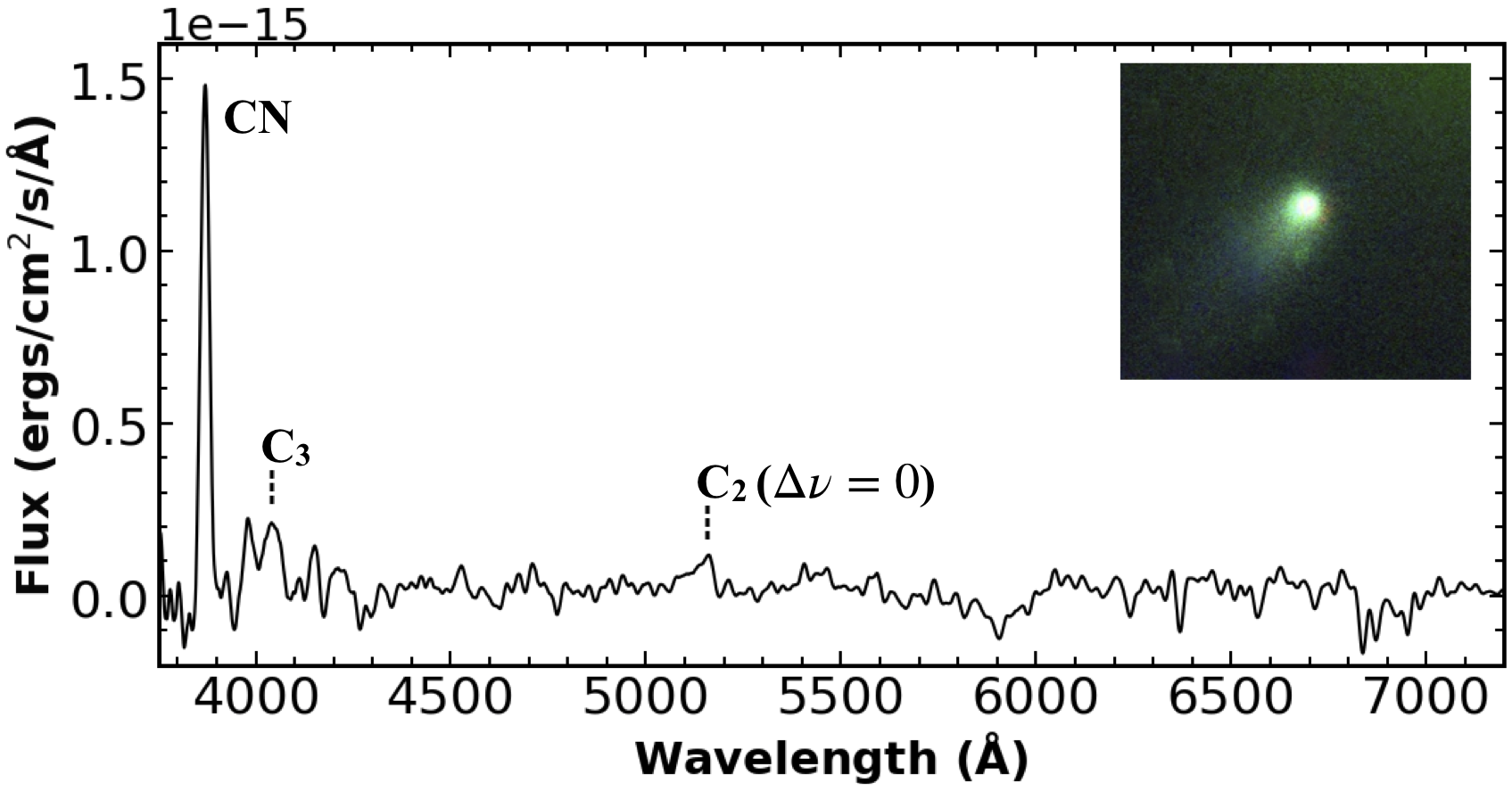}
   \caption{Optical spectrum of 2I/Borisov observed with the HFOSC instrument on HCT on 2019-12-22.96 UT.  Inset shows the RGB colour composite view of the comet on the same night using images taken with HFOSC. }
              \label{borisov}
    \end{figure*}
  
\subsection{Himalayan Chandra Telescope (HCT)}

The  2~m HCT is located at Hanle, Ladakh (Longitude: $78\degr~57\arcmin~49.8\arcsec$ East; Latitude : $32\degr~46\arcmin~46.3\arcsec$ Altitude : $4475$~m).  We used the Himalaya Faint Object Spectrograph and Camera (HFOSC) on the HCT to observe the comet 2I/Borisov. HFOSC uses a 2K $\times$ 4K CCD having a pixel size of 15 $\times$ 15 microns with a CCD pixel scale of $0.296\arcsec$ per pixel. 2K $\times$ 2K portion of the CCD is used during imaging and 1.5K $\times$ 4K during spectroscopy in order to get a good spatial coverage for the comet. \\
Spectroscopic observations were carried out on 30$^{th}$ November and 22$^{nd}$ December, using HFOSC with grism 7 providing a wavelength range of 3700 - 7200 \AA. A long slit, $11\arcmin$ in length and $1.92\arcsec$ in width, was used for the observation of the comet and another long slit $15.41\arcsec$ in width, was used for the observation of spectroscopic standard star.  Both the slits are placed horizontally in the E-W direction. With this configuration, we have a spectral resolving power of 1330 for comet observations. The comet was tracked at non-sidereal rate during spectroscopic and imaging exposures, using the keystone mode available in the HCT telescope control system. In this mode a trackfile containing the comet's  altitude-azimuth coordinates at regular intervals, is given as an input to the system. Exposures of 1800 seconds were obtained on the photocenter of the comet in spectroscopic mode.  Separate sky frame was not obtained due to time constraint. Standard star, \textsc{HD74721} (A0V type), from the catalog of spectroscopic standards in IRAF\footnote{The full list of spectroscopic standards available in IRAF can
be found in \url{http://stsdas.stsci.edu/cgi-bin/gethelp.cgi?onedstds}} was observed for flux calibration. Halogen lamp spectra, zero exposure frames and FeAr lamp spectra were obtained for flat fielding, bias subtraction and wavelength calibration respectively. 
\\Imaging observations were also carried out during the same epochs using the  Johnson-Cousins \textit{BVRI} filters. Multiple frames were obtained for each epoch, with exposure varying over the range 120-300 s. Ru 149 photometric field was also observed in all above mentioned filters in order to perform photometric calibration of the comet images. Twilight flats were recorded and bias frames were also taken at regular intervals during the night to correct the pixel to pixel response and remove the bias offset respectively. 

\subsection{Mount Abu InfraRed Observatory (MIRO)}
The Mount Abu InfraRed Observatory (MIRO) is located at Mount Abu, Rajasthan (Longitude : $72\degr~46\arcmin~47.5\arcsec$ East;  Latitude: $24\degr~39\arcmin~8.8\arcsec$ North; Altitude : $1680~$m).  One of the backend instruments available is a $1024 \times 1024$ pixel EMCCD camera (Ixon) from Andor.    The EMCCD is mounted at the Cassegrain focal plane.  With $2\times2$ on-chip-binning mode we achieve a plate scale of 0.36 arcsec/pixel.  
A CFW-2-7 (7 position, 2 inch diameter per filter) model filter wheel, from Finger Lakes Instrumentation, holds the Johnson-Cousins  {\citep[$UBVRI$, as described by][]{bessell}} broadband filters.
Imaging observation in the \textit{BVRI} filters were carried out on 24$^{th}$, 25$^{th}$ and 27$^{th}$ of December 2019. Again, Ru 149 field was chosen as a standard star field to be used for photometric calibration.  Twilight flats were obtained in all filters to normalise the pixel to pixel response of the CCD. The non-sidereal track mode built into the in-house developed telescope control software was used for comet observations.\\
\noindent

\begin{table*}
\centering
\setlength\tabcolsep{8pt}
\renewcommand{\arraystretch}{1.5}
\captionsetup{justification=centering}
\caption{{Observational Log}}
\begin{tabular}{|c|r|c|r|r|c|r|r|}
\hline 
\hline
\Tstrut\Bstrut  & & \multicolumn{1}{|c|}{\textbf{Telescope}} & \multicolumn{1}{|c|}{\textbf{Heliocentric}}   & \multicolumn{1}{|c|}{\textbf{Geocentric}} & \multicolumn{1}{|c|}{\textbf{Distance scale}} & \multicolumn{1}{|c|}{\textbf{Phase}}  &  \\
\multicolumn{1}{|c|}{\textbf{Date}} &  \multicolumn{1}{|c|}{\textbf{Time}} & \multicolumn{1}{|c|}{\textbf{Facility}} &\multicolumn{1}{|c|}{\textbf{Distance} (r$_{H}$)} & \multicolumn{1}{|c|}{\textbf{Distance} ($\Delta$)} & \multicolumn{1}{|c|}{\textbf{at photo-centre}} & \multicolumn{1}{|c|}{\textbf{angle}}    & \multicolumn{1}{|c|}{\textbf{Airmass}} \\
\multicolumn{1}{|c|}{[UT]}  & \multicolumn{1}{|c|}{[UT]} &  &  \multicolumn{1}{|c|}{[AU]}  & \multicolumn{1}{|c|}{[AU]} &    \multicolumn{1}{|c|}{[Km/arcsecond]} &  \multicolumn{1}{|c|}{[$^\circ$]} &   \\ \hline
\Tstrut\Bstrut 30/11/2019 & 23.04 &HCT & 2.013 & 2.049 & 1486 & 28.08& 1.83\\
\Tstrut\Bstrut 22/12/2019 & 23.16 &HCT & 2.031 & 1.94 & 1407 &28.56 & 2.6\\
\Tstrut\Bstrut 24/12/2019 & 22.25  &MIRO & 2.039 & 1.938 & 1406 & 28.5 & 1.96\\
\Tstrut\Bstrut 25/12/2019 & 21.62 &MIRO & 2.043 & 1.937 & 1405 & 28.45 & 2.23\\
\Tstrut\Bstrut 27/12/2019 & 21.84 &MIRO & 2.05 & 1.936 & 1404 & 28.36 & 2.1\\
\hline
\end{tabular}

\label{observations_1}
\end{table*}

\section{Data reduction and analysis}
\label{sec:2}
Careful reduction techniques are necessary to  reduce, calibrate and extract information from the raw data. The following sub-sections discuss, in brief, the various steps used in the process of analysing the data obtained from spectroscopy and imaging.

\subsection{Spectroscopy}\label{spec}
Standard \textsc{IRAF} routines were used to reduce the spectroscopic observational data. The bias files taken throughout the night were median combined, flat-field images were average combined and normalised in order to perform the basic reductions of comet and standard frames. Owing to the fact that the observatory is at a very high altitude and the comet was being observed for very long exposures, the raw files are contaminated by a large number of cosmic rays. The Laplacian Cosmic Ray Identification \citep{LAcosmic}\footnote{\url{http://www.astro.yale.edu/dokkum/lacosmic/}} package was added to \textsc{IRAF} and was used to remove most of the cosmic rays present in the spectroscopic raw data. \textsc{IRAF's \textit{apall}} module was used to extract the 1D spectrum from the comet, calibration lamp and standard star frames. For both epochs, an aperture of $17.76\arcsec$ (corresponding to 60 pixels, centered on the comet) was used to extract the comet spectrum. 
The corresponding physical distance at the photo-centre can be estimated using the distance scale column from Table \ref{observations_1}.
The sky spectrum required for subtraction was extracted using a similar aperture about 60\arcsec~away from the photo-centre, free from cometary emissions (see Appendix \ref{appsky}  for details). The spectrum for each aperture used was extracted by tracing along the dispersion axis.  The lines in the FeAr calibration lamp spectrum were identified using the \textit{identify} task and the solution was used to wavelength calibrate the comet and standard star spectra.  Using the \textit{standard} and \textit{sensfun} tasks, an instrument sensitivity function was derived by comparing the observed standard star flux with the catalog values available in \textsc{IRAF}. The resulting instrument sensitivity function was used to flux calibrate the comet spectrum. Extinction correction was also applied along with the flux calibration. Details regarding the continuum subtraction using solar spectrum is explained in Appendix \ref{app:cont}. A flux calibrated spectrum of the comet 2I/Borisov is shown in Figure \ref{borisov}, clearly indicating the presence of emissions from CN, C$_2$ and C$_3$ with the latter two highly depleted, similar to what has been observed by \cite{opitom_borisov}, \cite{lin_borisov}, \cite{kareta_borisov}, \cite{leon_borisov} and \cite{borisov_muse}. The inset image represents the RGB view of the comet using the same instrument. 
\\The area under the curve within the wavelength range covering the full emission bands of different molecules, as mentioned in \cite{langland-shula}, was used to obtain the total flux of the observed emissions. The uncertainties in the flux were obtained from the noise in the parts of the spectrum adjacent to the emission bands. The total number of molecules (N) present in the aperture used was calculated using

\begin{equation}
\label{mol_tot}
    N  =\frac{4\pi\Delta^2}{g}\times F,
\end{equation}
where $g$ is the fluorescence efficiency, $\Delta$ is the geocentric distance and $F$ is the total flux inside the aperture used for extraction. The total number of molecules present in the aperture was converted into the number present in the entire coma with the help of the Haser factor (further described in Appendix \ref{factor}). The Haser factor for both the epochs were obtained from Schleicher's website\footnote{\url{https://asteroid.lowell.edu/comet/}} for an aperture radius of 8.88\arcsec~(corresponding to 30 pixel, half the total aperture used). Since the Haser factor is derived for a circular aperture, the obtained value was adjusted for our aperture area of $1.92\arcsec \times 17.76\arcsec$. The total number of molecules was then divided by the lifetime of the daughter molecule ($\tau$) to obtain the production rate. The values for $g$ and $\tau$ at 1 AU were taken from \cite{Ahearn_85}. \citet{schleicher_CN_2010} have tabulated the g-factor of CN for different heliocentric distances and velocities, which we have used in our calculations. The scale lengths were scaled by $r_h^2$ and the fluorescence efficiency of other molecules by $r_h^{-2}$ in order to obtain the appropriate values to be used at the corresponding heliocentric distance ($r_h$).
\subsection{Imaging}
\label{imaging_reduction}
A self scripted Python routine was used to perform all basic data reduction techniques (bias subtraction, flat fielding and median combining) on both comet and standard star images. Aperture photometry was performed on the comet and standard stars (Ru 149, Ru 149B, Ru 149D, Ru 149E) using the \textsc{PHOTUTILS} package in Astropy. The instrumental magnitudes of the standard stars were then corrected for both extinction and color. Extinction coefficient values (magnitude/airmass) of various filters, as given in \cite{hct_extinction}, were used to apply the extinction correction to the instrumental magnitudes of the standard stars as well as the comet images obtained from HCT. The coefficients for Mount Abu were taken from an in house project carried out for computing the extinction values for the site. The zero point offset in magnitude was then computed with the help of Landolt's standard star magnitudes as given in \cite{landolt}. The comet instrumental magnitude were then corrected for zero point to obtain its apparent magnitude in various filters. These are tabulated  in Table \ref{borisov_mag}.

\begin{figure}
   \centering
   \includegraphics[width=\linewidth]{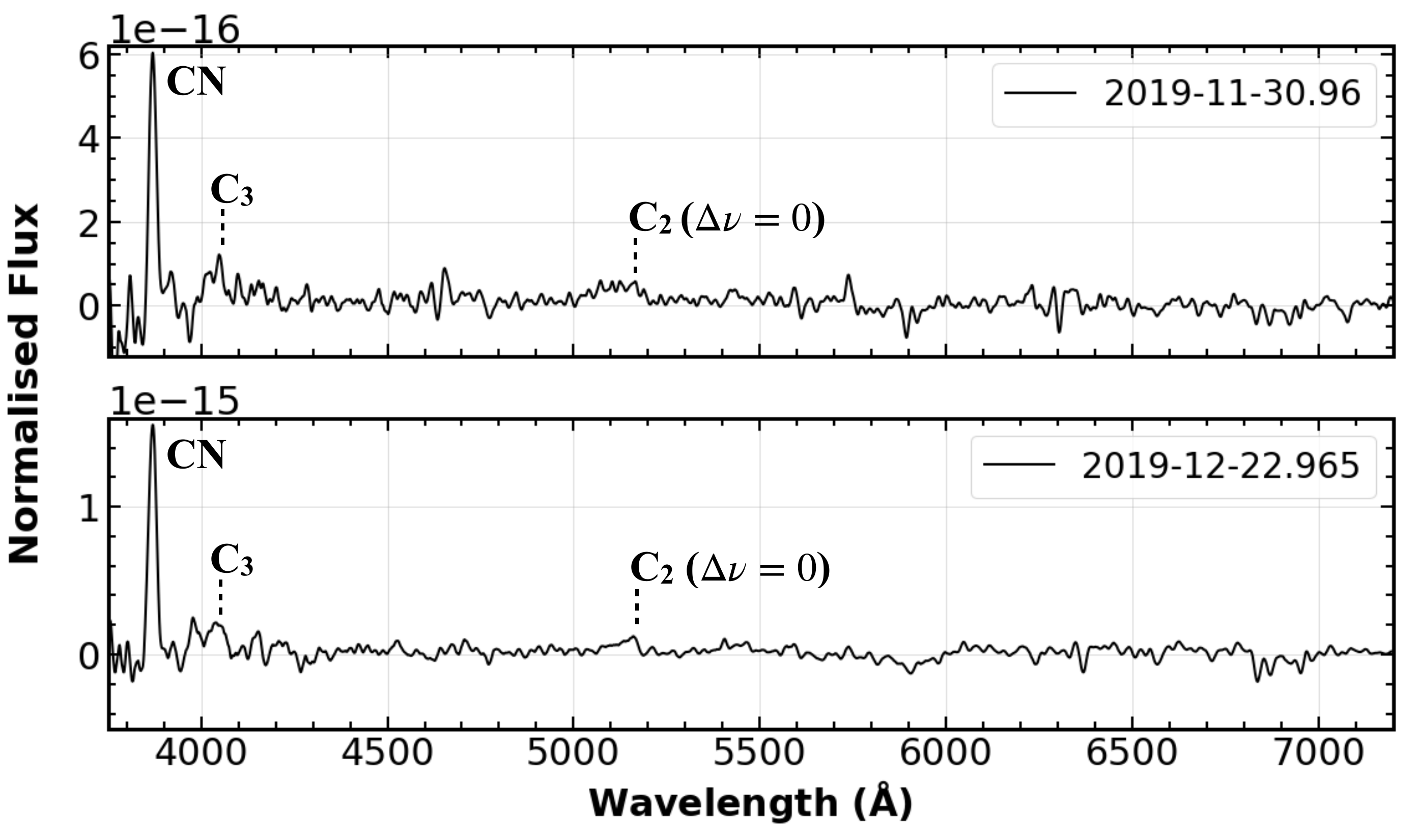}
   \caption{Optical spectra of 2I/Borisov as observed on the pre and post perihelion epochs.}
              \label{borisov_datecomp}
    \end{figure}

\section{Results \& discussion}
\label{sec:3}
\subsection{Spectroscopy}
\begin{figure*}
   \centering
   \includegraphics[width=0.8\textwidth]{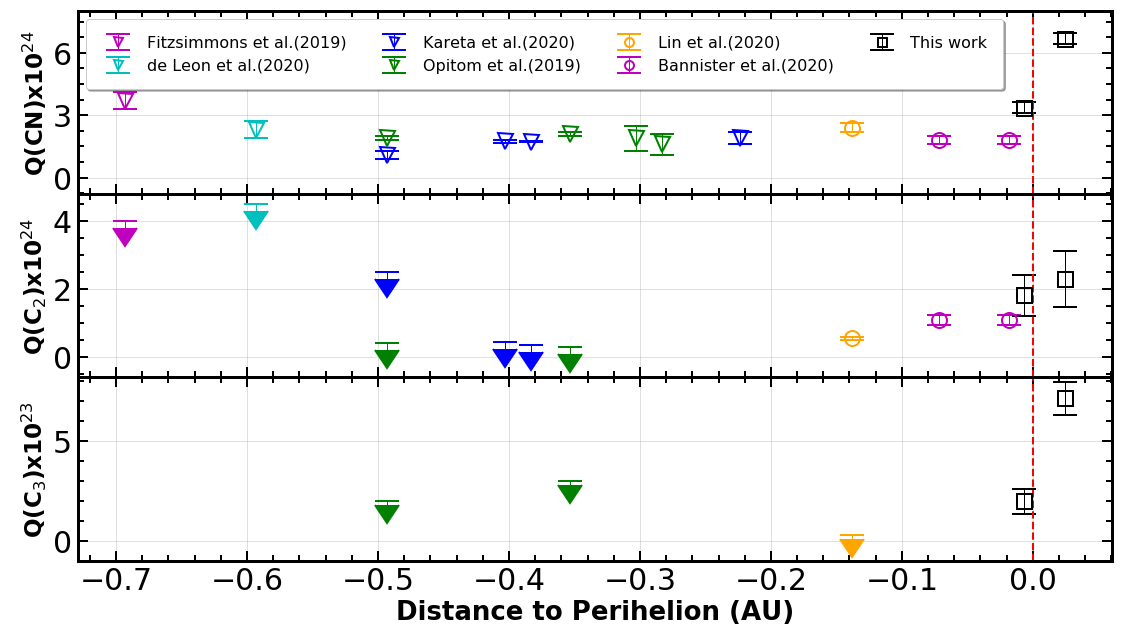}
   \caption{Comparison of pre-perihelion production rates (measurements/upper limits) of CN (upper panel), C$_2$ (middle panel) and C$_3$ (lower panel) reported in \protect\cite{CN_detected},
   \protect\cite{opitom_borisov},
   \protect\cite{kareta_borisov},
   \protect\cite{leon_borisov},
   \protect\cite{lin_borisov} and
   \protect\cite{borisov_muse} with the pre and post perihelion production rates of same molecules as computed in this work.}
   \label{borisov_comp_CNC2C3}
    \end{figure*}

\begin{figure}
   \centering
   \includegraphics[width=\linewidth]{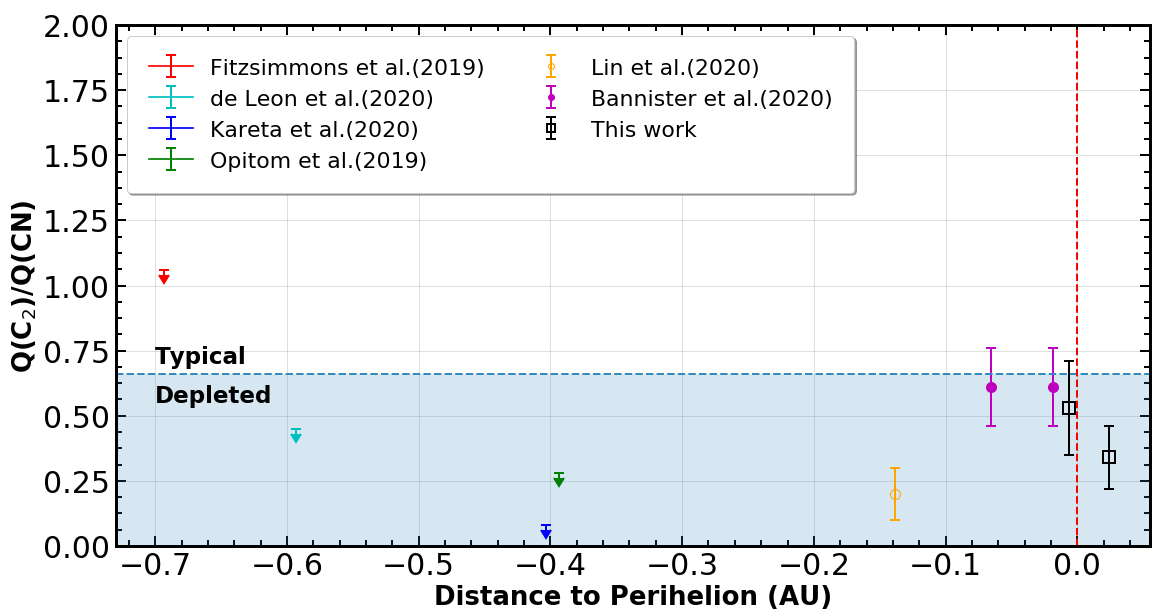}
   \caption{Cumulative comparison of the pre and post perihelion production rate ratio of C$_2$ and CN of 2I/Borisov observed in the current work with the values reported by \protect\cite{CN_detected}, \protect\cite{opitom_borisov}, \protect\cite{kareta_borisov}, \protect\cite{leon_borisov}, \protect\cite{lin_borisov} and \protect\cite{borisov_muse} while the comet was in an in-bound orbit. The shaded region represents the area of carbon chain depleted comets in our Solar system, for which Q(C$_2$)/Q(CN) < 0.66, as defined by \protect\cite{Ahearn_85}. The red dashed vertical line represents the perihelion of the comet on 8$^{th}$ December 2019 with q$\sim$2.0066 AU.}
              \label{borisov_comp_all}
    \end{figure}

From successful spectroscopic observations of the interstellar visitor, carried out on two epochs, pre and post perihelion,  we detect the presence of CN radical, C$_2$ ($\Delta \nu$ = 0) Swan band and C$_3$ emission (latter two being highly depleted) as shown in Figure \ref{borisov_datecomp}. The production rates of CN, C$_2$ and C$_3$ computed for both epochs, as mentioned in Section \ref{spec}, are listed in Table \ref{borisov_prod}. The comet has been monitored in spectroscopy by various groups with heliocentric distance ranging from 2.7 to 2.02 AU. The current work reports the production rates for the comet pre and post perihelion and hence contributes a valuable data point towards studying the characteristics of the emissions in the comet post perihelion.\\
The production rate of CN, C$_2$ and C$_3$, reported in this work, pre-perihelion, is comparable with the values reported in other observations, as shown in Figure \ref{borisov_comp_CNC2C3}, with slight increase in the rate which can be accounted by the fact that the comet was approaching perihelion. Among the clear detection of C$_2$ as reported by \cite{lin_borisov}, \cite{borisov_muse} and the current work, an increasing trend in the production rate can be observed, while it is difficult to compare the same with the upper limits reported in the other observations. Adding an important data point to the spectroscopic observation of the rare visit of the interstellar comet, the production rate of CN and C$_3$ shows drastic change with only a slight variation in that of C$_2$, post-perihelion.
The production rate ratio, Q(C$_2$)/Q(CN), was computed for both the epochs and the comet was seen to be depleted in carbon chain molecules, according to the classification criterion defined by \cite{Ahearn_85}. Also, the comet can be classified as depleted in carbon chain molecules according to the criterion defined by \cite{cochran_30years}, where the production rates of both C$_3$ and C$_2$ with respect to CN are considered. Figure \ref{borisov_comp_all} compiles the values of Q(C$_2$)/Q(CN) as reported from all the other observations with our own observations. It is interesting to observe that there is an increase in the production rate ratio with heliocentric distance (until perihelion), which is not common among Solar system comets for a minimal change in heliocentric distance \citep{Ahearn_85, cochran_30years}. Along the orbit of the comet, the behaviour has changed from  highly depleted to a moderately depleted comet, as it approached perihelion. Even though the production rate ratio, reported in our work, pre-perihelion, Q(C$_2$)/Q(CN) = 0.54,  is comparable to the values reported by \cite{borisov_muse}, it is surprising to observe that the value has dropped to Q(C$_2$)/Q(CN) = 0.34,  post-perihelion, once again making the comet highly depleted in carbon chain molecules. \cite{langland-shula} have reported the variation in production rate ratio (Q(C$_2$)/Q(CN)) with increasing heliocentric distance in a sample of Solar system comets. However,  \cite{Ahearn_85},  \cite{cochran_1992} and \cite{cochran_30years}   did not observe any  variation in the production rate ratios for a \textit{minimal} change in the heliocentric distance. In the current work, even though the production rate ratios are comparable within the errors, there is an indication of a possible asymmetry post-perihelion (see Figure \ref{borisov_comp_all}).  However, this cannot be confirmed with the limited post-perihelion data currently available. 

As shown in Figure \ref{borisov_comp_CNC2C3}, we also notice an asymmetry in the production rates of CN and C$_3$ post-perihelion. Generally, such asymmetries in production rates are observed among the short period comets of our Solar system \citep[eg.][]{opitom_67P, Ahearn_2P}, close to perihelion.  This asymmetry is expected either due to the illumination of different areas of the nucleus having different surface processing during their orbit or due to the presence of a less volatile surface, depleted in most of the molecules. Once these layers get  disintegrated by the solar radiation, the less depleted surface of the comet gets exposed resulting in an increased flux in emissions. However, \citet{borisov_COrich} and \citet{borisov_cordiner_CO}  report that 2I/Borisov has an extremely high abundance of carbon monoxide, implying that the surface of the comet has not undergone a sufficiently intense heat processing to cause the depletion of the top volatile surface. In addition, such high CO abundance is usually uncommon among the short period comets \citep{Dello_Russo_CO_SPC}.  All these reported facts and the observed asymmetry in production rates around perihelion,  makes us raise a question on the chemical homogeneity of the material present in the nucleus of 2I/Borisov or the difference in the volatile nature of the molecules present in the comet nucleus. \cite{borisov_muse} suggests a possibility of heterogeneous composition in the comet based on the observed increase in C$_2$ activity close to perihelion. Based on the observed very high abundance of CO in 2I/Borisov, \cite{borisov_COrich} points out the possibility of the comet having formed beyond the CO ice line of its parent stellar system.  Since the results from the current work are also in agreement with the suggestions regarding the heterogeneity, it is possible that the comet was formed in a stellar system beyond the CO ice line undergoing a very inhomogeneous mixing of various volatile compounds present in the proto-stellar nebula.  

\cite{Ahearn_water_C2} states that, since the parent molecules of C$_2$ are primarily contained in the grains of H$_2$O ice, the production of C$_2$ is directly related to the activity in the icy grains of H$_2$O, while production of CN and C$_3$ are not. \cite{Combi_C2_CHON} discuss the possibility of $C_2$ being produced from a primary parent molecule frozen in the icy mix of the nucleus and also directly from CHON grains at temperatures $\sim 500~K$. In the current scenario where the perihelion distance of the comet is 2.0066 AU, the temperature from solar radiation would not be high enough for CHON grains to be a primary source for C$_2$. Also, the influence of CHON grains would result in a flattening of the spatial profile of C$_2$ as per the CHON grain halo (CGH) model  proposed by \cite{Combi_C2_CHON}. Such a spatial flattening has not been reported yet in the case of 2I/Borisov. On the other hand, \cite{borisov_waterprod} reports that the water production in 2I/Borisov had increased drastically from November to December, close to perihelion and then decreased rapidly by December 21$^{st}$. With the contribution from CHON grains being ruled out, only the activity in H$_2$O ice can explain the increase in production rate of C$_2$ close to perihelion and hence the initial increase in Q(C$_2$)/Q(CN). The drop in the ratio post perihelion is due to an increased activity of CN while C$_2$ activity had not changed substantially. Results from our work also support the possibility reported by \cite{borisov_muse}, regarding the heterogeneity in the comet nucleus. This would have resulted in a fresh layer, rich in  carbon chain parent molecules trapped in the icy grains, being exposed and hence resulting in the steep increase of C$_2$ production rate  along with the water production rate. We also infer a possibility that, as the southern hemisphere of the nucleus was illuminated, after perihelion, a fresh unexposed area of the comet started sublimating as proposed by \cite{borisov_kim}. This resulted in the drastic increase in production rates of CN and C$_3$ with only a minimal change in the C$_2$ production rate owing to the reduced water production rate. Even though \cite{AHearn_CN_CHON} and \cite{Fray_CN} discusses the prospect of CHON grains being a possible parent source of CN, it can be ruled out in this case since the contribution would be very less due to the perihelion distance of the comet as discussed earlier. 
We also observe an abrupt discontinuity in the production rate of CN as compared to that for C$_2$ (see Figure \ref{borisov_comp_CNC2C3}).  This strongly suggests that both of them come primarily from different sources as discussed by \cite{Ahearn_water_C2}. This behaviour can also be considered as a confirmation, in this work, that the parent molecules of both CN and C$_3$ resides mainly in the comet nucleus whereas that of C$_2$ is mostly present in the icy grains  of the coma. The difference in activity of C$_3$ and C$_2$ also confirms that the parent molecules of these emissions should be entirely different as mentioned in \cite{CNC2C3_yamamoto}. \\

\begin{table*}
\centering
\renewcommand{\arraystretch}{1.5}
\setlength\tabcolsep{5pt}
\caption{{Gas production rates of comet 2I/Borisov at different heliocentric distances}}
\begin{tabular}{|r|c|r|r|r|r|r|c|c|}
\hline
\hline
 \multicolumn{1}{|c|}{\textbf{Date}} &  \multicolumn{1}{|c|}{\textbf{Exposure}} &\multicolumn{1}{|c|}{\textbf{r{$_{H}$}}}   & \multicolumn{1}{|c|}{$\Delta$} & \multicolumn{3}{|c|}{\textbf{Production Rate} (molec/sec)}  & \multicolumn{1}{|c|}{\textbf{Production rate ratio}}&\multicolumn{1}{|c|}{\textbf{Dust to gas ratio}}\\ 
\cline{5-7}
 \multicolumn{1}{|c|}{[UT]} & \multicolumn{1}{|c|}{[s]}&   \multicolumn{1}{|c|}{[AU]} & \multicolumn{1}{|c|}{[AU]} & \multicolumn{1}{|c|}{\textbf{CN}} & \multicolumn{1}{|c|}{\textbf{C{$_{2}$}}({$\Delta \nu = 0$})} & \multicolumn{1}{|c|}{\textbf{C{$_{3}$}}} & \multicolumn{1}{|c|}{\textbf{Q(C$_2$)/Q(CN)}} & \multicolumn{1}{|c|}{\textbf{log[($Af\rho$)$_R$/Q(CN)]}}\\ 
  & &  &  & \multicolumn{1}{|c|}{$\times 10^{24}$}  & \multicolumn{1}{|c|}{$\times 10^{24}$}  &\multicolumn{1}{|c|}{$\times 10^{23}$}  &  \\
\hline  
 2019-11-30.96 & 1800 & 2.013 & 2.049 & 3.36 $\pm$ 0.25 & 1.82 $\pm$ 0.60 & 1.97 $\pm$ 0.52 & 0.54 $\pm$ 0.18 & -22.24 $\pm$ 0.12  \\
 2019-12-22.965 & 1800 &  2.031 & 1.94 & 6.68 $\pm$ 0.27 & 2.30 $\pm$ 0.82 & 7.14 $\pm$ 0.74 & 0.34 $\pm$ 0.12 & -22.57 $\pm$ 0.12\\ 
\hline 
\end{tabular}
\label{borisov_prod}
\end{table*} 

\begin{table*}
\centering
\setlength\tabcolsep{4pt}
\renewcommand{\arraystretch}{1.5}
%\captionsetup{justification=centering}
\caption{{Apparent magnitude (m), absolute magnitude (H), effective scattering cross section (C$_e$) and $Af\rho$ computed for various observational epochs}}
\begin{tabular}{|c|c|c|c|c|c|c|r|r|r|r|}
 \hline 
 \hline
%\Tstrut\Bstrut 
\multicolumn{1}{|c|}{\textbf{Date}} & \multicolumn{4}{|c|}{\textbf{m$^a$}} &
    \multirow{2}{*}{\textbf{H}$_R^b$}
 & \multirow{2}{*}{ \textbf{C}$_e$[Km$^2$]}&
\multicolumn{4}{|c|}{\textbf{$Af\rho$ [cm]} } \\ 
\cline{2-5}\cline{8-11}
  \multicolumn{1}{|c|}{[UT]} & \multicolumn{1}{|c|}{\textit{B}}& \multicolumn{1}{|c|}{\textit{V}}& \multicolumn{1}{|c|}{\textit{R}}& \multicolumn{1}{|c|}{\textit{I}}&  & & \multicolumn{1}{|c|}{\textit{B}}& \multicolumn{1}{|c|}{\textit{V}}& \multicolumn{1}{|c|}{\textit{R}}& \multicolumn{1}{|c|}{\textit{I}}\\

 \hline 
 2019-11-30 & 17.41 $\pm$ 0.08 & 16.59 $\pm$ 0.07 & 16.16 $\pm$ 0.10 & 15.64 $\pm$ 0.06 & 12.16 $\pm$ 0.07 & 157 $\pm$ 12 & 107 $\pm$ 3 & 110 $\pm$ 3 & 120 $\pm$ 4 & 138 $\pm$ 5 \\
 2019-12-22 & 17.62 $\pm$ 0.08 & 16.86 $\pm$ 0.07 & 16.31 $\pm$ 0.09 & 15.79 $\pm$ 0.06 & 12.39 $\pm$ 0.06 & 127 $\pm$ 9 & 80 $\pm$ 2 & 87 $\pm$ 3 & 97 $\pm$ 3 & 124 $\pm$ 5 \\
 2019-12-24 & $-$ & $-$ & 16.37 $\pm$ 0.14 & 15.81 $\pm$ 0.07 & 12.46 $\pm$ 0.10 & 118$\pm$ 13 & \multicolumn{1}{|c|}{$-$} & \multicolumn{1}{|c|}{$-$} & 94 $\pm$ 2 & 120 $\pm$ 3 \\
 2019-12-25 & 17.72 $\pm$ 0.22 & 16.9 $\pm$ 0.17 & 16.41 $\pm$ 0.15 & $-$ & 12.56 $\pm$ 0.11 & 108 $\pm$ 12 &  66 $\pm$ 5 & 86 $\pm$ 2 & 95 $\pm$ 2 & \multicolumn{1}{|c|}{$-$}\\ 
 2019-12-27 & $-$ & $-$ & 16.44 $\pm$ 0.14 & $-$ & 12.58 $\pm$ 0.10 & 106 $\pm$ 11 & \multicolumn{1}{|c|}{$-$} & \multicolumn{1}{|c|}{$-$} & 94 $\pm$ 2 & \multicolumn{1}{|c|}{$-$}\\
\hline
\multicolumn{11}{|l|}{$^a$ An aperture size of 10,000 Km has been used on all epochs for all filters to compute the magnitude}\\
\multicolumn{11}{|l|}{$^b$Absolute magnitude computed using Eq.\ref{abs_mag} from the corresponding apparent magnitude in R filter}\\
\hline
 \end{tabular}
 \label{borisov_mag}
\end{table*}

\subsection{Imaging}

The comet was observed in imaging mode on 5 epochs (Table. \ref{observations_1}) in the Bessells's \textit{BVRI} filters from both HCT and MIRO. The images were reduced and apparent magnitudes were computed as explained in section \ref{imaging_reduction}.

\subsubsection{Optical colors and $Af\rho$}\label{afrho}
From the magnitudes computed, as given in Table \ref{borisov_mag}, the optical colours of the comet were found to be; \textit{(B-V)} = 0.80 $\pm$ 0.05, \textit{(V-R)} = 0.49 $\pm$ 0.04, \textit{(R-I)} = 0.53 $\pm$ 0.03 \textit{(B-R)} = 1.29 $\pm$ 0.06. These colours are in good agreement with the colours reported by \cite{borisov_initial}, which are slightly redder than the solar colours \citep{sun_color} and similar to that of 1I/'Oumuamua \citep{oumuamua_Jewitt}. The colours of the interstellar comet are also surprisingly similar to the mean colours of long period comets in the Solar system \citep{comet_color_jewitt}. The \textit{(B-V)}, \textit{(V-R)} and  \textit{(R-I)} colours, after transformation to the SDSS photometric system as described in \cite{sdss2JC}, also compares within uncertainties in measurement to the colours reported by \cite{bolin_borisov_imaging} and \cite{hui_borisov}. The available magnitudes were also used to compute  $Af\rho$, a proxy to the amount of dust produced \citep{Ahearn_Bowel_slope}. The obtained values of $Af\rho$ (see Table \ref{borisov_mag}), for 30$^{th}$ November and 22$^{nd}$ December, in V band is found to be similar to the values reported by \cite{borisov_waterprod}, for the same wavelength band, during nearby epochs (1$^{st}$ December and 21$^{st}$ December respectively). The equation for average slope of the curve of reflectivity, as mentioned in \cite{Ahearn_Bowel_slope}, when used with the observed magnitudes provides a slope $S^\prime = (9.9 \pm 1.2)\%/10^3$~\AA~for the red-end  (6400-7900 \AA) and $S^\prime = (13.5 \pm 1.5)\%/10^3$~\AA~for the blue-end (4200 - 5500 \AA). The observed average slope at the red end is consistent with the values reported by \cite{lin_borisov} [$S^\prime = 9.2\%/10^3$\AA], \cite{leon_borisov} [$S^\prime = (10 \pm 1)\%/10^3$\AA], \cite{kareta_borisov} [$S^\prime = 11\%/10^3$\AA] and \cite{hui_borisov} [$S^\prime = (10.6 \pm 1.4)\%/10^3$\AA]. These values of spectral slope of 2I/Borisov suggests that the dust composition present in the cometary coma could be similar to those observed in the D-type asteroids \citep{active_asteroids}, a suggestion first proposed by \citet{borisov_Dtype} from their spectroscopic observations of 2I/Borisov. The observed slope at the blue end cannot be compared with the values reported through spectroscopy, since the magnitudes measured using the broad band filters, \textit{B \& V}, would be largely affected by the emissions from CN and C$_2$ respectively.
The dust-gas ratio, as shown in Table \ref{borisov_prod} is also similar to the dust-gas ratio of carbon chain depleted Solar system comets \citep{Ahearn_85}.  These are clear indications of the similarity in dust composition of 2I/Borisov with Solar system comets implying a high possibility of the comet formation process similar to Solar system happening in other stellar systems. 

\subsubsection{Absolute magnitude and effective scattering cross section}\label{abs_ce}
\begin{figure}
    \centering
    \includegraphics[width=\linewidth]{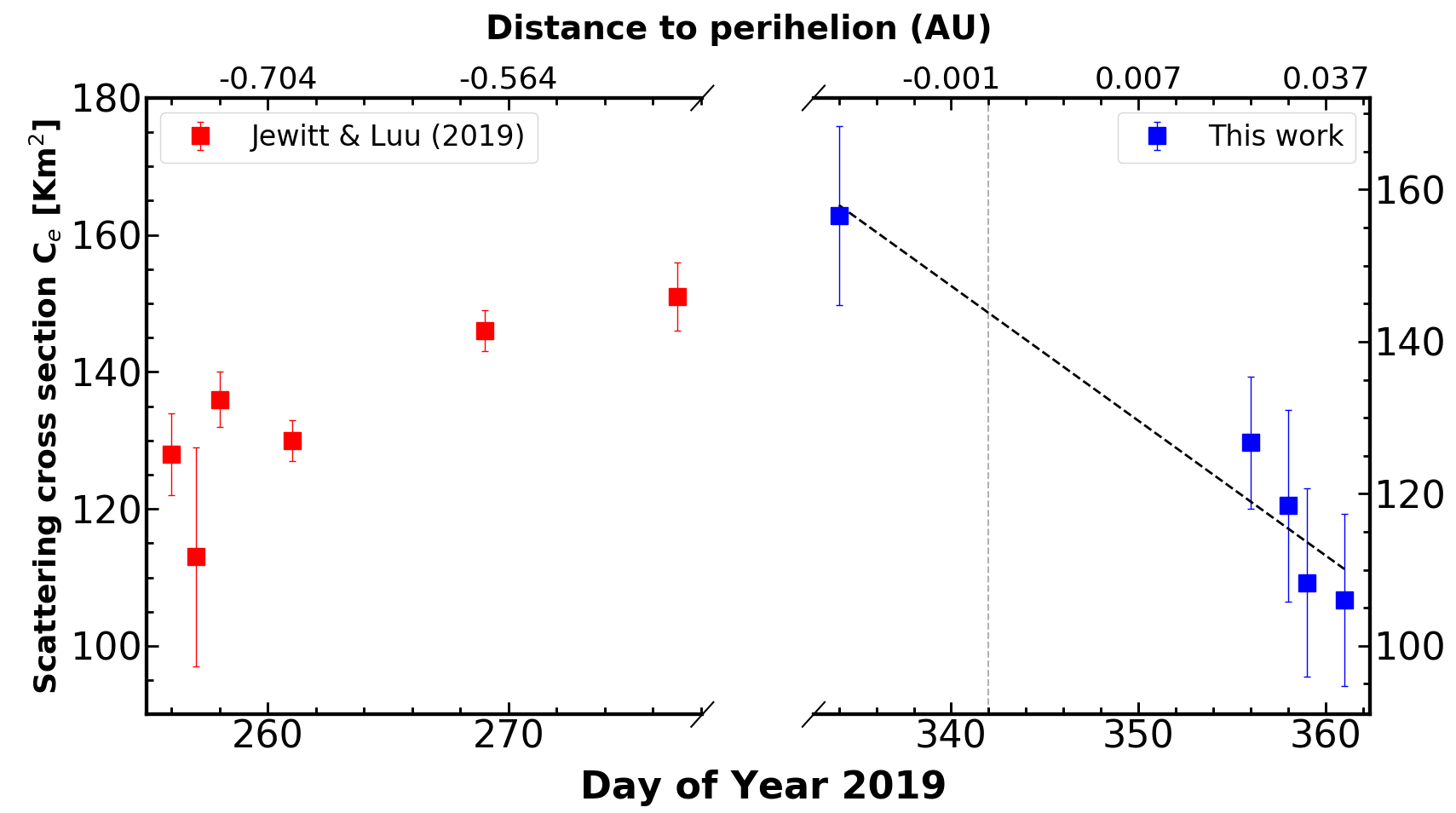}
    \caption{Variation of scattering cross section, computed using equation \ref{ce:eq} as a function of time, expressed as Day of Year 2019, with  Day 1 being January 1 2019. The top-axis is labelled with the distance to perihelion.
    The black dashed line is the best linear least-squares fit having a gradient -1.77 $\pm$ 0.22 Km$^{2}$/day. The vertical grey dashed line represents the perihelion of the comet. 
    The red points are the scattering cross section as reported by \citet{borisov_initial} while the comet was approaching 
    perihelion.}
                \label{borisov_scattering}
    \end{figure}

The apparent magnitude is a function of the heliocentric distance, geocentric distance and the phase angle at the time of observation. Hence, the absolute magnitude (H), which corresponds to the magnitude of the comet at a heliocentric and geocentric distance of 1 AU and a phase angle of 0$^\circ$, given by

\begin{equation}\label{abs_mag}
H = m - 5log(r_H \Delta) + 2.5log[\phi(\alpha)],
\end{equation}
was computed. Here $m$ is the apparent magnitude in the respective filter, $r_H$ is the heliocentric distance, $\Delta$ is the geocentric distance and $\phi(\alpha)$ is the phase function\footnote{Composite Dust Phase Function for Comets \url{https://asteroid.lowell.edu/comet/dustphaseHM_table.txt}} corresponding to the phase angle at the time of observation, as defined in \cite{schleicher_phasefunction}. The R band absolute magnitude can be used to compute the effective scattering cross section in order to investigate the nature of activity in the comet. The effective scattering cross section (C$_e$) is computed using the following equation;
\begin{equation}
 C_e = \frac{\pi r_0^2}{p}10^{0.4[m_{\odot,R}-H_R]},
\label{ce:eq}
\end{equation}
where $r_0$ is the mean Earth-Sun distance in Km, $p$ is the geometric albedo of the cometary dust and $m_{\odot,R}$ is the solar apparent magnitude in the R band. Using the values of $r_0 = 1.5 \times 10^8$ Km and $m_{\odot,R} = -26.97$ from \cite{sun_mag}, the above equation reduces to $C_e = (1.15\times10^6/p)\times10^{-(0.4H_R)}.$
For this work, the albedo (p) of the comet was chosen as 0.1, typical for comet dust \citep{albedo_zubko}, as used in \cite{borisov_initial}, \cite{hui_borisov} and \cite{bolin_borisov_imaging}. Figure \ref{borisov_scattering} depicts the decreasing trend in the scattering cross section as a function of days in the year 2019. The grey dashed line at 342$^{nd}$ day represents the perihelion of the comet (8/12/2019) and the solid dashed line represents a linear least-squares fit, having a best fit slope ${d(C_e)/dt}=-1.77~\pm~0.22$ Km$^2$ d$^{-1}$. A mean nuclear radius (r) can be computed for
$C_e = 123~Km^2$ as $r \leq 3.1$~Km. This value, in close agreement with the sizes reported by other groups \citep{CN_detected, leon_borisov, borisov_initial}, can only be considered as an upper limit since the photometric aperture used is highly influenced by the dust in the coma. For comparison, the variation in scattering cross section, when the comet was in bound, reported by \cite{borisov_initial}, is also included in Figure \ref{borisov_scattering}, where an increasing trend is observed. As per the variation of cross section with heliocentric distance reported by \cite{bolin_borisov_imaging},  the cross section is seen to be increasing till the day when 2I crossed the water-ice line at 2.5 AU and decreasing later on.  Clubbing the short range trends reported in \cite{borisov_initial} (before 2I crossed the  water-ice line) and in this work (after 2I crossed the water-ice line) along with the larger range trend reported in \cite{bolin_borisov_imaging},  it is clear that there was a steady increase in the scattering cross section till the water-ice line beyond which it decreased systematically. This observation is not in agreement with the variation in scattering cross section reported by \cite{hui_borisov}, where the cross section is seen to be continuously reducing.  
\\
Further, making use of the rate of change in effective cross section, the rate of dust production can be calculated as,
\begin{equation}
    \frac{dM}{dt} = \frac{4}{3}\rho \Bar{a} \frac{d(C_e)}{dt},
\end{equation}
where $\rho$ is the particle density, $\Bar{a}$ is the mean particle size. In this work we have accepted the values of $\rho = 1 g/cm^3$ \citep{borisov_initial} and $\Bar{a}=100~\mu m$ \citep{borisov_initial, hui_borisov}. Substituting these values, we get the average net mass loss rate $dM/dt = -2.74 \pm 0.34~kg~s^{-1}$. This value depicts the rate of change in dust mass over the observed period. The value is negative since the dust produced from the comet is not able to compensate for the dust lost from the photometric aperture ($\rho \sim 10^4$ Km), implying that the absolute amount of dust production is reducing over the time period, resulting in the comet getting fainter.

\subsubsection{Connecting Af$\rho$, dust production rate and sublimation flux}
As mentioned in \cite{dust_model_borisov}, the following equation,
\begin{equation}
    Af\rho = \frac{3AQ_d}{\rho v_d s_0},
\end{equation}
where $A$ is the geometric albedo, $v_d$ is the dust ejection velocity, $\rho$ is the particle density and s$_0$ is the average particle size,
can be used to convert the observed $Af\rho$ (in meter) (see Section \ref{afrho}) into the dust production rate (Q$_d$). Using the same values for the parameters, as used in the previous section, we get a relation,
$Q_d = 3.3v_d(Af\rho).$
Accepting the value of dust ejection velocity as $v_d\sim8$ m s$^{-1}$ \citep[for epochs close to perihelion as reported by ][]{hui_borisov}, using Af$\rho$ values for V band, we obtain a dust production rate, $Q_d \sim 30$~kg~s$^{-1}$.  This absolute value of dust production rate is in agreement to the values reported in \cite{dust_model_borisov}, \cite{borisov_outburst_jewitt} and \citet{borisov_kim}.  Considering this dust to be produced by water-ice sublimation, the patch of area supplying this dust can be computed as $A = Q_d/f_s$, where $f_s$ is the specific rate of mass sublimation flux at equilibrium.  According to \cite{sublimation_jewitt}, the specific rate of mass sublimation flux at equilibrium, $f_s~(Kg~m^{-2}~s^{-1})$, for a body at a heliocentric distance R is obtained from the equation, 
\begin{equation}
\frac{F_\odot (1-A)}{R^2} cos\theta = \epsilon \sigma T^4 + L(T)f_s ,
\label{sublimation}
\end{equation}
where $F_\odot$ is the solar constant, $A$ is the albedo, $\epsilon$ is the emissivity and $L(T)$ is the latent heat of sublimation of ice at temperature T. For the comet 2I/Borisov at 2.7 AU, 
 \cite{borisov_initial} obtains $f_s = 4 \times 10^{-5}$ kg m$^{-2}$ s$^{-1}$. Considering the change in temperature from 2.7 to 2.013 AU to be only $\Delta T~\sim 30 K$  \citep{comet_temperature}, the value for latent heat of sublimation of water ice does not change significantly. Hence from Equation. \ref{sublimation}, for the current work, we obtain the specific rate of mass sublimation flux to be $f_s = 7.2~\times~10^{-5}$  kg m$^{-2}$ s$^{-1}$. Inserting this value in the above defined relation for area, provides $A = 0.4$ Km$^2$, which is equal to the surface area of a sphere of radius $r = 0.18$ Km. The computed nuclear radius is in good agreement with the lower limits reported in observations using the Neil Gehrels-\textsc{Swift} Observatory’s Ultraviolet/Optical Telescope \citep{borisov_waterprod} and the Hubble Space Telescope \citep{borisov_nucleus_jewitt}.\\
Assuming that the  empirical relation mentioned in \cite{visual2waterprod} holds for this interstellar comet, we can compute the water production rate from the visual \textit{V} band magnitude (reduced to a geocentric distance of 1 AU). Using the observed reduced magnitudes, the expected water production rates (Q(H$_2$O)) for 30$^{th}$ November and 22$^{nd}$ December 2019 are $(9.7 \pm 0.9) \times 10^{26}$ molec/s and $(7 \pm 0.8) \times 10^{26}$ molec/s respectively. Even though the expected water production rate for 30$^{th}$ November is consistent, within uncertainties, with the observed rate reported by \cite{borisov_waterprod} for 1$^{st}$ December, their reported rate on 21$^{st}$ December is much lower than expected from our observations for 22$^{nd}$ December. Emphasising on the fact that the empirical relation cannot be used to get precise measurements of water production rates, but only for order of magnitude estimates, we would like to point out that, such an unexpected drastic decrease in water production rate may be due to the heterogeneous composition of the nucleus, as discussed earlier, along with the low abundance of H$_2$O, uncommon in Solar system comets, as reported in \cite{borisov_COrich}.
  
\subsection{Conclusions}
In this work we present the optical spectroscopic and imaging observations of the interstellar comet 2I/Borisov, before and after perihelion, using the 2-m HCT, Hanle and MIRO, 1.2~m Mt.Abu telescopes. Spectroscopic study shows clear emissions from CN, C$_2$ and C$_3$ pre and post perihelion and detects a drop in production rate ratio, Q(C$_2$)/Q(CN) post perihelion.
Imaging study reveals a systematically reducing $Af\rho$ and effective cross section, using which a possible size range of the nucleus has been computed. Using these observational results, we arrive at the following conclusions:
 \begin{enumerate}
      \item The computation of production rates of molecules CN, C$_2$ and C$_3$ shows an increase, comparable to the observations by other groups, as the comet approached perihelion with an asymmetry in the emission observed post perihelion. 
      \item The low value of the production rate ratio, Q(C$_2$)/Q(CN), implies that the comet is depleted in carbon-chain molecules. The ratio had increased as the comet moved closer to perihelion, making it a moderately depleted one, with a later decrease in the ratio after perihelion passage.
      \item We infer a chemical heterogeneity in the comet surface due to which there was a drastic surge in the C$_2$ emissions as the comet approached perihelion. We also infer that an initially unexposed surface of the comet would have been exposed to solar radiation post perihelion resulting in the substantial increase in the production rates of CN and C$_3$. 
      \item The $Af\rho$ values computed from the imaging observations, and hence the dust-to-gas ratio, are consistent with the numbers observed for Solar system comets, depleted in carbon-chain molecules. This may be an indication that the parent stellar system of 2I/Borisov would have undergone a formation process somewhat similar to that of the Solar system.
      \item The optical colours of the comet, \textit{(B-V)} = 0.80 $\pm$ 0.05, \textit{(V-R)} = 0.49 $\pm$ 0.04, \textit{(R-I)} = 0.53 $\pm$ 0.03 \textit{(B-R)} = 1.29 $\pm$ 0.06, are slightly redder than Solar and similar to the mean colour of large number of comets of our Solar system.
      \item The water production rate computed using an empirical formula considering the V band magnitude, was found to match the rate reported before perihelion, but failed to match the rate reported after perihelion, due to a drastic drop in the observed water production rate.  This maybe due to the fact that the empirical relation (only to get  an order of magnitude estimate) is defined for Solar system comets  and not applicable to 2I/Borisov, an interstellar comet with very low abundance in H$_2$O.
      \item The possible size of the nucleus was deduced to be $0.18 \leq r \leq 3.1$ Km. This range is in very good agreement with the sizes reported by other groups.
      \item Considering all the observational evidences, we infer that the comet 2I/Borisov was formed in a proto-stellar system undergoing a very inhomogeneous mixing of various volatile compounds beyond the CO ice line.
\end{enumerate}

\section*{Acknowledgements}
We thank the referee for the valuable comments and suggestions which have improved the manuscript.
We acknowledge the local staff at the Mount Abu InfraRed Observatory for their help. We thank the staff of IAO, Hanle and CREST, Hoskote, that made these observations possible. The facilities at IAO and CREST are operated by the Indian Institute of Astrophysics, Bangalore. 
Work at PRL is supported by the Dept of Space, Govt. of India.

%%%%%%%%%%%%%%%%%%%%%%%%%%%%%%%%%%%%%%%%%%%%%%%%%%
\section*{Data Availability}

The data underlying this article will be shared on reasonable request to the corresponding author. 



\bibliographystyle{mnras}
\bibliography{reference} % if your bibtex file is called example.bib

%\begin{thebibliography}{99}
%\end{thebibliography}

%%%%%%%%%%%%%%%%%%%%%%%%%%%%%%%%%%%%%%%%%%%%%%%%%%

%%%%%%%%%%%%%%%%% APPENDICES %%%%%%%%%%%%%%%%%%%%%

\appendix
\section{Sky subtraction}
\label{appsky}
\begin{figure}
    \centering
    \includegraphics[width=1\linewidth]{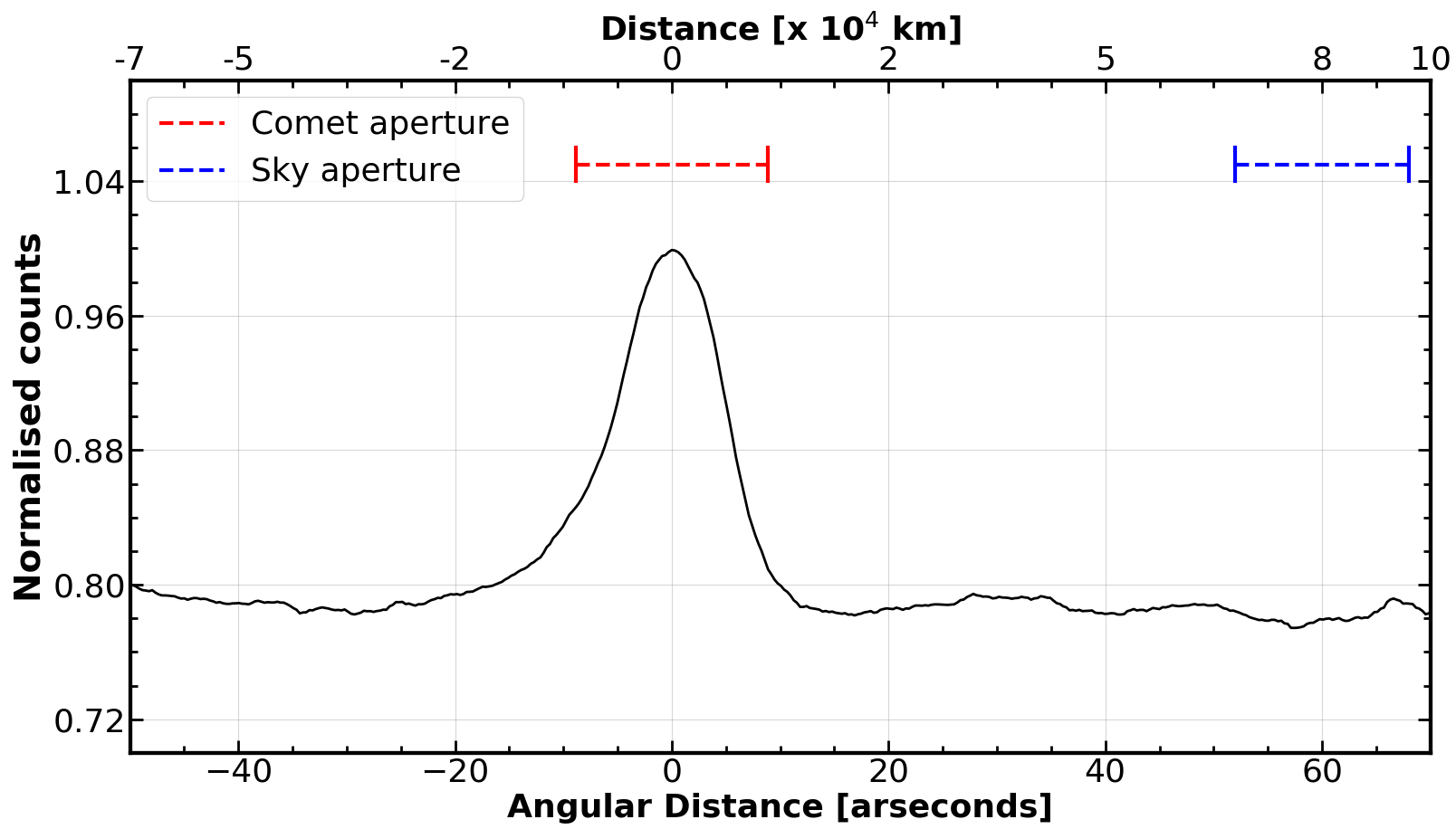}
    \caption{Profile along the slit (spatial direction) showing the locations of the apertures used for extracting the flux from the comet and the background sky.}
                \label{spec_radial}
    \end{figure}

\begin{figure}
    \centering
    \includegraphics[width=1\linewidth]{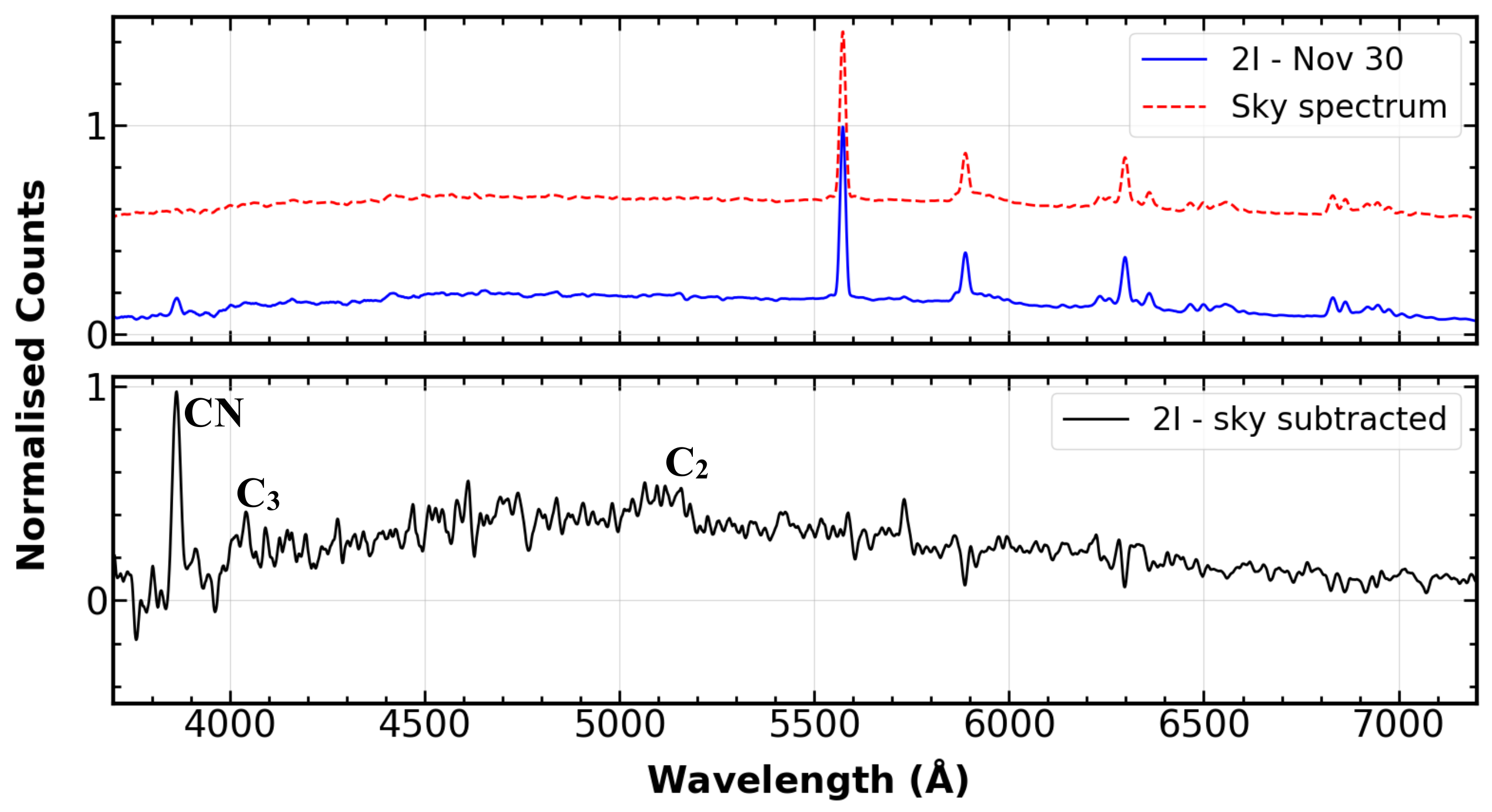}
    \caption{\textit{Top panel} : Comet spectrum (blue solid line) and sky spectrum (red dashed line, with a constant offset) illustrating the presence of strong atmospheric telluric lines . \textit{Bottom panel} : Comet spectrum after sky correction with the detected emissions marked. }
                \label{skycor}
    \end{figure}

During the observations for this work, a separate sky frame was not obtained due to time constraints. On analysing the profile along the slit, in the spatial axis, it appears that 2I/Borisov has an extent of about 20\arcsec.~
Taking this into account, we used an equal sized aperture for the comet and the sky about 60\arcsec~apart.  
Figure \ref{spec_radial} depicts the positions of both these apertures over-plotted on the profile along the slit, in the spatial axis. The normalised wavelength calibrated spectra of both comet and sky, with a constant offset
in the sky spectrum is shown in the top panel of Figure \ref{skycor}. The bottom panel depicts the corresponding sky corrected spectrum, with the detected emissions marked.

\section{continuum correction}
\label{app:cont}
\begin{figure}
    \centering
    \includegraphics[width=1\linewidth]{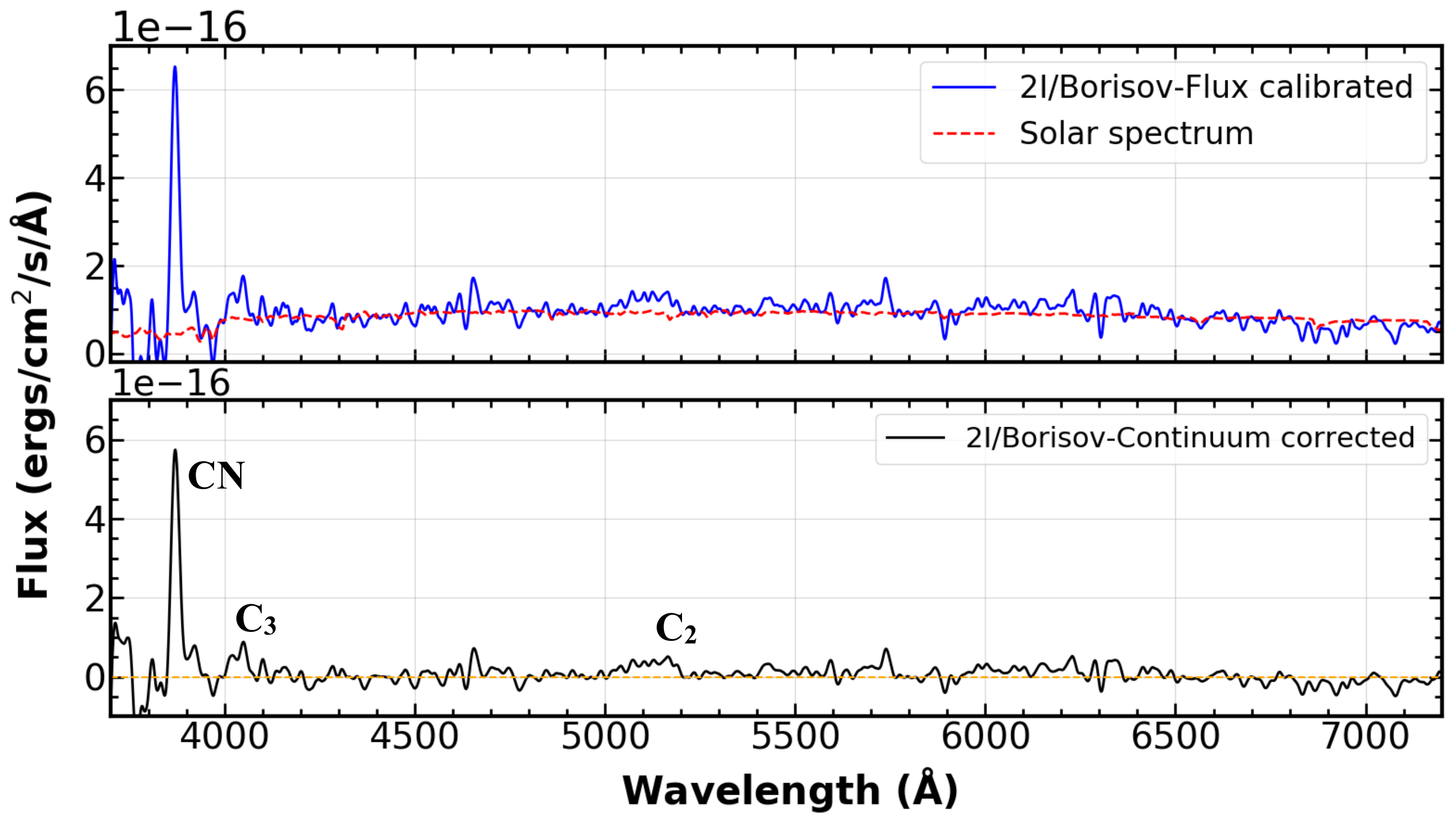}
    \caption{\textit{Top panel} : Comet spectrum (blue solid line) overplotted with a standard solar spectrum (red dashed line) scaled to the continuum level. \textit{Bottom panel} : Comet spectrum after continuum correction with the detected emissions marked. }
                \label{contcor}
    \end{figure}

In order to extract the total flux of the various molecules from the cometary emission spectrum it is necessary to remove the contribution from the continuum. For this, either a solar analog star is observed along with the comet observations or a standard solar spectrum is used. In the present case, we have used a standard solar spectrum, scaled, re-sampled to match the resolution of the instrument and corrected for slope to take into account the redder nature of the comet dust (see Figure \ref{contcor}). 

\section{Haser factor}
\label{factor}

During long slit spectroscopic observations of comets only a part of the total coma is being observed. Hence, the observed flux of each molecular species is to be extrapolated to obtain the total flux of the same in the entire coma of the comet. Haser model \citep{haser}, provides a factor called the Haser factor which is the ratio of total number of molecules present in the aperture used to the total number of molecules present in the whole coma. The reciprocal of this factor, the Haser correction, can be used to extrapolate the observed flux so as to estimate the molecular abundance in the entire coma. Since the Haser model assumes a spherically symmetric coma with uniform outflow of gas, we compute the Haser factor for a circular aperture by making use of the radius of the aperture (in arcseconds). This factor can be computed using the web calculator implemented by Prof. Schleicher on his website(\url{https://asteroid.lowell.edu/comet/}).  In the current work, we have used a rectangular slit ($1.92 \arcsec \times 17.76 \arcsec$, 60 pixel along the slit with comet at the centre).  Hence, we have chosen an aperture radius of $8.88\arcsec$ (corresponding to 30 pixel) in order to compute the Haser factor. Since the computed Haser factor is for a circular aperture, the factor was normalised to unit area and then multiplied by the area of the rectangular aperture used, so as to obtain the factor required for this work.  

% If you want to present additional material which would interrupt the flow of the main paper,
% it can be placed in an Appendix which appears after the list of references.

%%%%%%%%%%%%%%%%%%%%%%%%%%%%%%%%%%%%%%%%%%%%%%%%%%

% Don't change these lines
\bsp	% typesetting comment
\label{lastpage}
\end{document}